\documentclass{nature}

\usepackage{graphicx}

\title{The nature of solar noise (SUBMITTED VERSION)}

\author{A.I. Shapiro$^{1}$, S.K. Solanki$^{1,2}$, N.A. Krivova${^1}$,  R. H. Cameron$^{1}$, K.L. Yeo${^1}$, W.K. Schmutz$^{3}$}

\def\aap{Astron. Astrophys.}

\def\apj{Astrophys. J.}

\def\solphys{Sol. Phys.}

\begin{document}

\maketitle

\begin{affiliations}
\item Max-Planck-Institut f{\"u}r Sonnensystemforschung, Justus-von-Liebig-Weg 3, 37077, G{\"o}ttingen, Germany
\item School of Space Research, Kyung Hee University, Yongin, Gyeonggi 446-701, Korea
\item Physikalisch-Meteorologishes Observatorium Davos, World Radiation Centre, 7260 Davos Dorf, Switzerland
\end{affiliations}

\begin{abstract}
The solar brightness varies on timescales from minutes to decades\cite{Claus_rev,Greg2016}. Determining the sources of such variations, often referred to as solar noise, is of importance for multiple reasons: a) it is the background that limits the detection of solar oscillations\cite{noise_limit}, b) variability in solar brightness is one of the drivers of the Earth's climate system\cite{TOSCA2012,MPS_AA}, c) it is a prototype of stellar variability\cite{Shapiro2014_stars,meunieretal2015} which is an important limiting factor for the detection of extra-solar planets. Here we show that recent progress in simulations and observations of the Sun makes it finally possible to pinpoint the source of the solar noise. We utilise high-cadence observations\cite{HMI_12,yeoetal2014} from the Solar Dynamic Observatory and the SATIRE\cite{krivovaetal2003,SATIRE} model to calculate the magnetically-driven variations of solar brightness. The brightness variations caused by  the constantly evolving cellular granulation pattern on the solar surface are computed with the MURAM\cite{MURAM} code. We find that surface magnetic field and granulation can together precisely explain solar noise on timescales from minutes to decades, i.e. ranging over more than six orders of magnitude in the period. This accounts for all timescales that have so far been resolved or covered by irradiance measurements. We demonstrate  that no other sources of variability are required to explain the data. 
Recent measurements of Sun-like stars by CoRoT\cite{COROT2} and Kepler\cite{KEPLER} uncovered brightness variations similar to that of the Sun but with much wider variety of patterns\cite{basrietal2013}.  Our finding that solar brightness variations can be replicated in detail with just two well-known sources will greatly simplify future modelling of existing CoRoT and Kepler as well as anticipated TESS\cite{TESS} and PLATO\cite{PLATO} data.
\end{abstract}
The advent of spaceborne measurements in 1978 led to the revelation that the Total Solar Irradiance (TSI, which is the spectrally integrated solar radiative flux at one au from the Sun)  varies on multiple timescales. 





The magnetic field has been named as the main source of solar irradiance variations on timescales of a day and longer\cite{TOSCA2012,MPS_AA}.  The largest concentrations of field at the solar surface are seen as dark sunspots, while ensembles of small concentrations form more diffuse bright faculae\cite{rev_mag}. 
The transits of magnetic features across the visible solar disk as the Sun rotates as well as their evolution cause quasi-periodic irradiance modulation on timescales of days to weeks, while changes in the overall magnetic activity level lead to irradiance variations on longer timescales. 

Models based on the concept that magnetic fields at the solar surface are responsible for irradiance variations successfully reproduce available measurements of solar brightness variations on timescales longer than a day\cite{TOSCA2012,MPS_AA}. At the same time the sources of variations on shorter timescales have until now remained unclear. These timescales  are typical of planetary transits across the disks of Sun-like stars.   Consequently, the absence of a clear picture of brightness variability even in the case of the Sun hinders assessing the critical limits put by stellar variability on the detectability of exoplanets.



Here we show that recent advances in observations and modelling allow reproducing TSI variability with high precision at all timescales from minutes to decades.  We compute the magnetic component of the TSI variability with the SATIRE-S model\cite{krivovaetal2003} (Spectral And Total Irradiance Reconstruction, with the suffix ``S'' denoting the ``satellite-era''), which is one of the most successful and refined models of magnetically-driven solar irradiance variability\cite{TOSCA2012,yeoetal2014}. The high-cadence  solar magnetograms and continuum images recorded by the Helioseismic and Magnetic Imager onboard the Solar Dynamics Observatory (SDO/HMI)\cite{HMI} allowed us to expand SATIRE-S for calculations of magnetically-driven TSI variability on timescales down to 12 minutes (see Methods section). Our calculations of granulation-driven TSI variability are based on recent 3D simulations\cite{beeck2} of  convective gas currents both, above and below the solar surface with the MURAM\cite{MURAM} code (see Methods section).

Some statistical models\cite{aigrain2004,Penza2009} of TSI variability account for larger convective structures, such as supergranules and mesogranules. However, currently there is no evidence that these structures have an intrinsic brightness contrast of non-magnetic origin\cite{rast2003} so that we do not include them in our modelling.  We also refrain from including oscillations (which dominate TSI variability at periods about 5 minutes and have been extensively used for helioseismology) in our modelling and focus on the TSI variations considered in helioseismology as noise. 

While the amplitude of the granulation component of the TSI variability does not depend on time, the magnetic component is linked to the specific configuration of  faculae and spots on the visible solar disk and thus depends on solar magnetic activity. Therefore we consider four intervals of the TSI record representing different levels of activity (Table~1). The three one-month intervals at 2-minute cadence allow studying the high-frequency component of the variability  (Fig.~1), whereas the 19-year interval with a daily cadence is used to assess long-term changes.

\begin{figure}
\resizebox{\hsize}{!}{\includegraphics{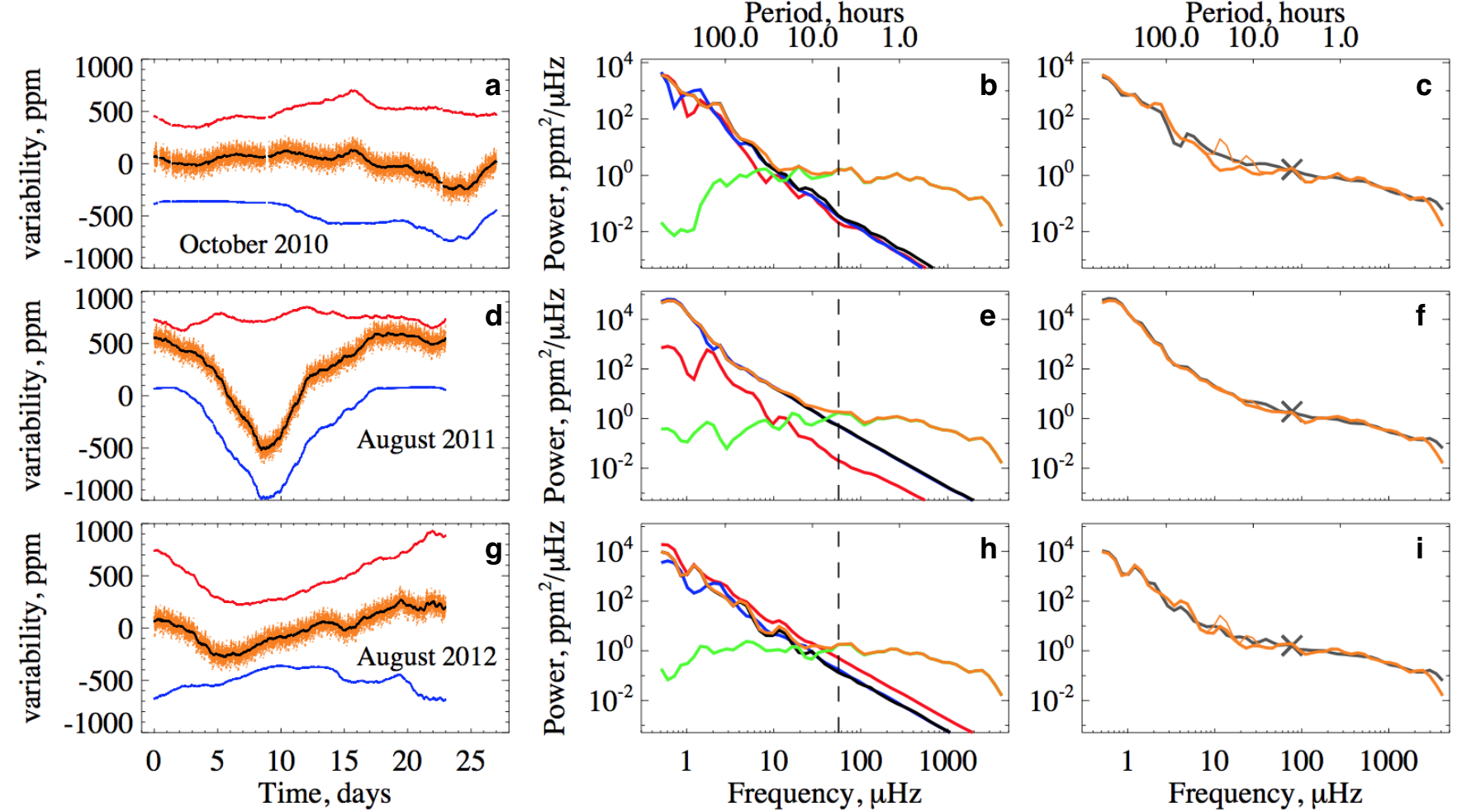}}
\caption{{\bf Short-term TSI variability at three intervals of very different activity level and variability of the Sun. a, d, g.} Calculated TSI variations (orange), the total magnetic (black), facular (red), and spot (blue) contributions to the TSI variation for three one-month intervals listed in Table~1. The plotted curves have been offset around zero for clarity. {\bf b, e, h.} Global wavelet power spectra of the calculated TSI variations. In addition to the variability components shown in a, d, g also the granulation components (green) are plotted. The vertical dashed lines indicate the period of 5 hours. {\bf c, f, i.} Global wavelet power spectra of the observed (gray) and modelled (orange) TSI variations. Plotted are the TSI variations calculated with and without correcting HMI data for the orbital velocity (thick and thin orange curves, respectively). The ``X'' symbol indicates the threshold between the PREMOS and VIRGO segments of the observed wavelet power spectra. The full description of  the observed and modelled datasets is given in the Methods section.}
\label{fig:mod_sp}
\end{figure}

\begin{table}
\caption{The intervals of the TSI evolution considered in this study.} 
\label{table:incl}
\centering 
\begin{tabular}{ c  | c  c  c  c} 
\hline
\hline
      Interval &   Period & Cadence  & Solar activity  \\ 
\hline 
  October 2010       &    04.10.2010 -- 31.10.2010   &  2 min  &    low   \\
  August 2011         &    23.07.2011 -- 16.08.2011   &    2 min   &   transit of a large sunspot group  \\
  August 2012         &    10.08.2012 -- 01.09.2012   &   2 min &   high facular, low spot coverage    \\
  Long         &    01.01.1996 -- 06.04.2015   &   1 day    &  most over cycles 23 and 24   \\
 \hline 
\end{tabular}
\end{table}
Our calculations show that the TSI variability on timescales of 5 hours (dashed lines in Fig. 1b, e, h) and shorter is entirely due to the granulation. The power spectrum is almost flat at periods between 1 and 5 hours (except statistical noise). For periods less than an hour the power decreases with frequency  due to the increasing coherence between granulation patterns. Since the period corresponding to the transition between the flat and decreasing parts of the power spectrum depends on the 
mean granule lifetime\cite{seleznyovetal2011}, the power spectra of stellar variations, observed with the Kepler and CoRoT  (and in future TESS and PLATO) missions, could provide a sensitive tool for determining lifetimes of stellar granules. 


Both, the granulation and the magnetic components of the TSI variability are important on timescales of 5 to 24 hours, while at longer periods the TSI variability is dominated by the magnetic field.
To better illustrate the role of faculae and spots 
we also separately plot the magnetically-driven variability due to spots and faculae.

It has been recently noticed\cite{Greg2016} that while TSI varies at a level of 200 ppm on timescales of minutes and exhibits significantly larger variations on timescales of days, transits of inner planets across the solar disk (in particular, the Mercury transit on May 9 2016 with a resulting TSI decrease of 50 ppm) could be clearly distinguished from solar noise in the TSI data. Our modelling allows explaining this curious observations: magnetically-driven TSI variations on timescales of planetary transits are small, while granulation-driven variations can be averaged out by smoothing TSI over time. For late-type stars the threshold timescale between granulation- and magnetically-driven brightness variations  is expected to be a function of magnetic activity (compare the crossings between the green and black lines in Figs. 1b, e, h), spectral class, and metallicity.

To test our calculations against observations we have utilised TSI time series obtained by the PREMOS instrument\cite{werner_PREMOS} onboard the PICARD mission and the VIRGO instrument\cite{VIRGO}  onboard the Solar and Heliospheric Observatory (SOHO). The PREMOS and VIRGO data are complementary in a sense that they allow reliable calculation of the  TSI power spectra on timescales longer and shorter than 3.5 hours, respectively (see Methods section). Consequently, we have created the  PREMOS/VIRGO composite TSI power spectra (see Methods section and Extended Data Figure 1) for the three one-month intervals considered in this study.

The model reproduces the power spectra of the TSI in  August 2011 and August 2012  remarkably well (Fig.~1f, i).
The deviation at timescales of about 5 minutes is due to the p-mode oscillations which are not included in our model. For the October 2010 interval, the agreement worsens somewhat  at timescales of about 10--20  hours, which might be a signature of overcorrection of the SDO/HMI data for the orbital velocity (see Methods section and Extended Data Figure 2).

\begin{figure}
\resizebox{\hsize}{!}{\includegraphics{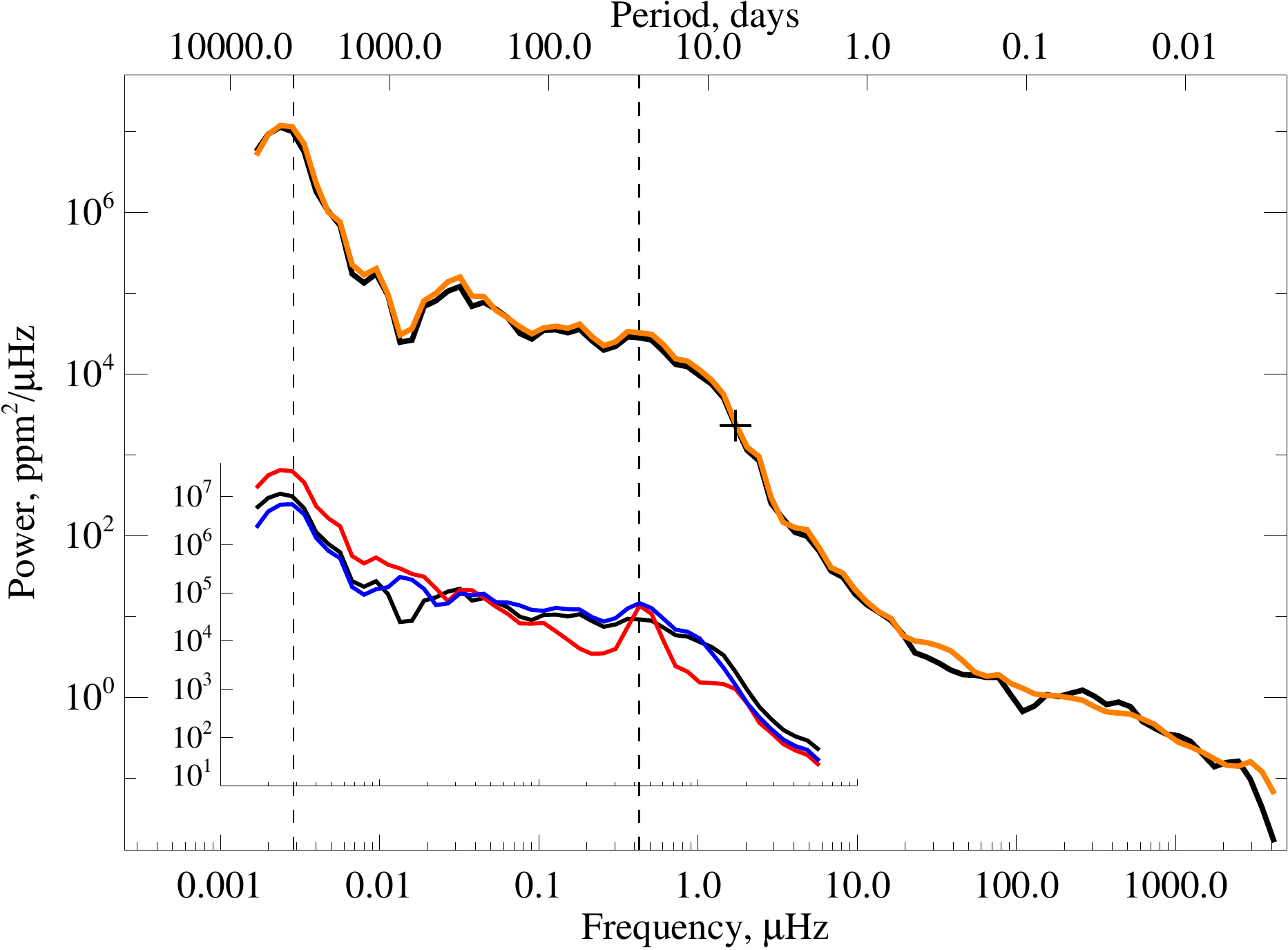}}
\caption{{\bf TSI variability on timescales from 4 minutes to 19 years.} Main panel, power spectra of modelled (black) and measured (orange) TSI variations. The plus sign indicates the threshold between power spectra for 1996-2015 and August 2011 time intervals (see Methods section for details). Inset, magnetically-driven TSI variations calculated with daily cadence (black) and their facular (red) and spot (blue) components. The vertical dashed lines  indicate 11-year and  27-day periods.}
\label{fig:mod_sp}
\end{figure}

{We have merged (Fig.~2) the power spectra calculated for the August 2011 (keeping its 4 minutes -- 7 days part) and ``Long''  (the 7 days -- 19 years part) intervals.} On the observational side we have merged corresponding parts of the PREMOS/VIRGO composite power spectrum for August 2011 and power spectrum calculated employing TSI values from the PMOD composite\cite{pmod_comp}, which is a de facto standard observational dataset of daily TSI time series. 

Our calculations accurately reproduce observed TSI variability on timescales covering more than 6 orders of magnitudes (4 minutes -- 19 years), over which the power spectral density changes by  8 orders of magnitudes (Fig.~2).  Thus a joint action of  granulation and magnetic field fully explains TSI variability (excluding the 5-minute oscillation component). The remarkable agreement between the model and the measurements cogently demonstrate that our understanding of these mechanisms, and the TSI variability in general, is fundamentally correct. 

Our model allows us to separately look at the magnetically-driven variability associated with spots and faculae. The transition between faculae- and spot-dominated regimes of the TSI variability occurs at a timescale of about one year (Fig.~2). Although magnetic TSI variations at timescales less than a year are primarily due to spots, the contribution of faculae to the TSI variability is still rather significant and comparable to that of spots at timescales of 2 to 7 days as well as at the solar rotational period, 27 days. The amplification of the facular signal at 2--7 days timescales is associated with the strong dependence of the contrast of faculae on their position on the visible solar disk\cite{shapiroetal2016}, while the peak at 27 days is attributed to long-lived large concentrations of faculae (whose lifetime can be up to several solar rotation periods).

Interestingly, the pronounced 27-day peaks in spot and facular components of TSI variability essentially cancel each other so that the total power spectrum  is almost flat around the 27-day period.  This implies that white light observations are not well suited for determining periods of stars similar to the Sun and explains the poor performance of standard methods of period determination for stars older than the Sun\cite{Aigrainetal2015}. 

The success of our model in replicating observed TSI variations with two well-understood components might be considered as a proof of concept for developing a similar model for Sun-like stars and explaining the data obtained by the existing and planned planet-hunting space missions\cite{COROT2,KEPLER,TESS,PLATO}. This is a feasible task since a) reliable simulations of granulation patterns on main sequence stars are gradually becoming available\cite{Ludwiget2009,beeck2}, b) new methods for extrapolating the magnetically-driven brightness variations from the Sun to Sun-like stars with various levels of magnetic activity observed at arbitrary angles between rotational axis and line-of-sight have been recently developed\cite{Shapiro2014_stars,shapiroetal2016}.


\newpage
\setcounter{figure}{0}
\renewcommand{\figurename}{Extended Data Figure}

\begin{methods}

\section*{High-cadence TSI data}\label{sub:data}
We have utilised TSI data obtained by the PMO6-V radiometer of the VIRGO instrument onboard the SOHO mission and by the PREMOS absolute radiometer onboard PICARD mission (Extended Data Fig.~1) to test our model. VIRGO records TSI with 1-minute cadence. Because of the failure of the shutter, the photon recording is not interrupted by the calibration procedure\cite{Claus1997}. This results in an advantage for this study, leading to a negligibly small high-frequency noise level of the VIRGO data\cite{FandL2004}. The drawback of the shutter failure is that VIRGO is unstable on timescales from about 3--5 to about 100 hours (one can clearly see the signature of the instrumental variations on such scales in the form of excess power in Extended Data Fig. 1b, c, d). The amplitude of the TSI variability recorded by VIRGO at these timescales is mainly attributed to the instrumental effects (e.g. it does not depend on solar activity level, see Extended Data Fig.~1f). PREMOS records TSI with 2-minute cadence, but unlike VIRGO it interrupts for calibration every minute. While individual PREMOS measurements are less precise than those of VIRGO, they do not suffer from the VIRGO calibration problems.   Building on the complementary character of the PREMOS and VIRGO data we have created composite power spectra for three one-month intervals, using VIRGO and PREMOS  on timescales below and above 3.5 hour, respectively (Extended Data Fig.~1.e, f, g).

\begin{figure}
\resizebox{0.65\hsize}{!}{\includegraphics{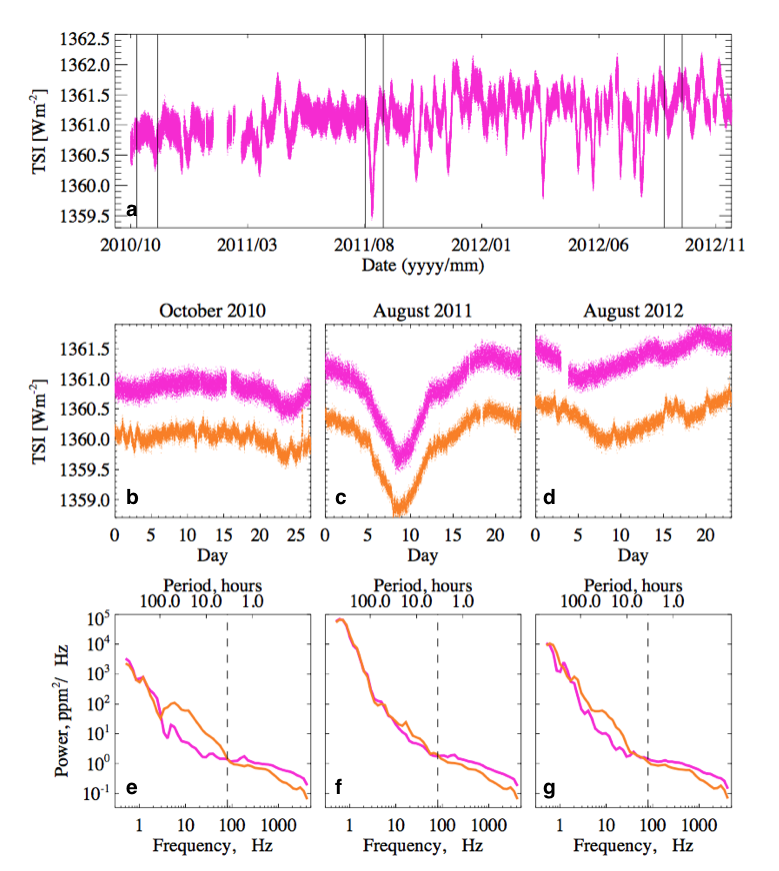}}
\caption{{\bf Spaceborne measurements of TSI variability. a.} TSI measured by PREMOS/PICARD between October 2010 and December 2012. Black vertical lines constrain three one-month intervals studied here (Table 1). {\bf b, c, d.} TSI measured by VIRGO (orange) and PREMOS (magenta) during the three intervals studied here. In all three panels day zero refers to the first day of the corresponding interval. {\bf e, f, g.} Global wavelet power spectra of the TSI variability measured by PREMOS (magenta) and VIRGO (orange) calculated for intervals shown in b, c, d. 
The vertical dashed lines indicate the period of 3.5 hours used as a threshold, with the PREMOS data being taken on the left (lower frequencies) and the VIRGO data on the right (high frequencies) to form the composite power spectra. }
\label{fig:mod_sp}
\end{figure}

\section*{SATIRE model}\label{sub:SAT}
The magnetically-driven TSI variability presented in this study is calculated with the SATIRE model. SATIRE decomposes the visible solar disk to magnetic features and the quiet Sun. The magnetic features are subdivided into three classes, faculae which encompass all bright magnetic features, sunspot umbrae, and sunspot penumbrae. The quiet Sun represents the part of the solar disk not covered by magnetic features. SATIRE calculates the full-disk solar brightness by weighting the fluxes from the magnetic features and the quiet Sun with their disk area coverages, taking into account  positions of the magnetic features on the solar disk. 

The spectra of the magnetic features and the quiet Sun have been pre-calculated\cite{sat_spectra} on a fine grid of wavelengths and disk positions with the ATLAS9 code\cite{kurucz1992,ATLAS9_CK}.  The source of the information about the disk area coverages and positions of magnetic features depends on the branch of the SATIRE model. For this study, we employ SATIRE-S, which is the most accurate implementation of SATIRE. SATIRE-S infers the disk area coverages of magnetic features and their positions from full-disk solar magnetograms and continuum images\cite{balletal2014, yeoetal2014}. 


\section*{12-minute SDO/HMI magnetograms}\label{sub:12}
Our modelling requires high-cadence information about the evolution of magnetic features on the visible solar disk.
Such information is available since May 2010 thanks to the SDO/HMI instrument. It simultaneously records continuum intensities and longitudinal magnetograms every 45 seconds. SDO/HMI data have been already used in conjunction with SATIRE-S for modelling solar variability on timescales longer than a day in the study by {\it Yeo et al. 2014}\cite{yeoetal2014}. 
We followed this approach but instead of producing  315 seconds averages as done by {\it Yeo et al. 2014}\cite{yeoetal2014} we utilised the 12-minute magnetograms and intensity images, which is less tedious and just as accurate.

To derive the disk area coverages of magnetic features, the 12-minute SDO/HMI magnetograms have been processed by the pipeline described in {\it Yeo et al. 2014}\cite{yeoetal2014} ,with one important exception. 
To avoid inconsistencies between SATIRE-S segments based on magnetograms from different instruments, {\it Yeo et al. 2014}\cite{yeoetal2014}  corrected SDO/HMI magnetograms for noise using the maps of the noise level of Michelson Doppler Imager  onboard the Solar and Heliospheric Observatory (SOHO/MDI). These maps were calculated with the {\it Ortiz et al. 2002}\cite{ortizetal2002} algorithm. In contrast, we directly applied the {\it Ortiz et al. 2002}\cite{ortizetal2002} algorithm to the SDO/HMI magnetograms. The noise level of SOHO/MDI is substantially higher than that of SDO/HMI so that the magnetograms reduced in this study contain more small-scale magnetic features with low magnetic flux than magnetograms reduced in {\it Yeo et al. 2014}\cite{yeoetal2014}.



The Doppler shift caused by the SDO orbital velocity\cite{HMI_12} triggers a noticeable 24 hours periodicity into the magnetic flux deduced from the HMI magnetograms and into the subsequently calculated  TSI contribution by faculae (Extended Data Fig.~2, blue lines). To take this into account we have calculated the mean signature of the 24 hours variations in the facular component for each of the three periods examined in this study (Extended Data Fig.~3) and subtracted it from the total magnetic and facular light curves (Extended Data Fig.~2, red lines).  Since this procedure might also remove part of the real TSI variability we plot both power spectra, those based on corrected and those based on uncorrected light curves and compare them to the measurements (Fig.~1).

\begin{figure}
\resizebox{0.75\hsize}{!}{\includegraphics{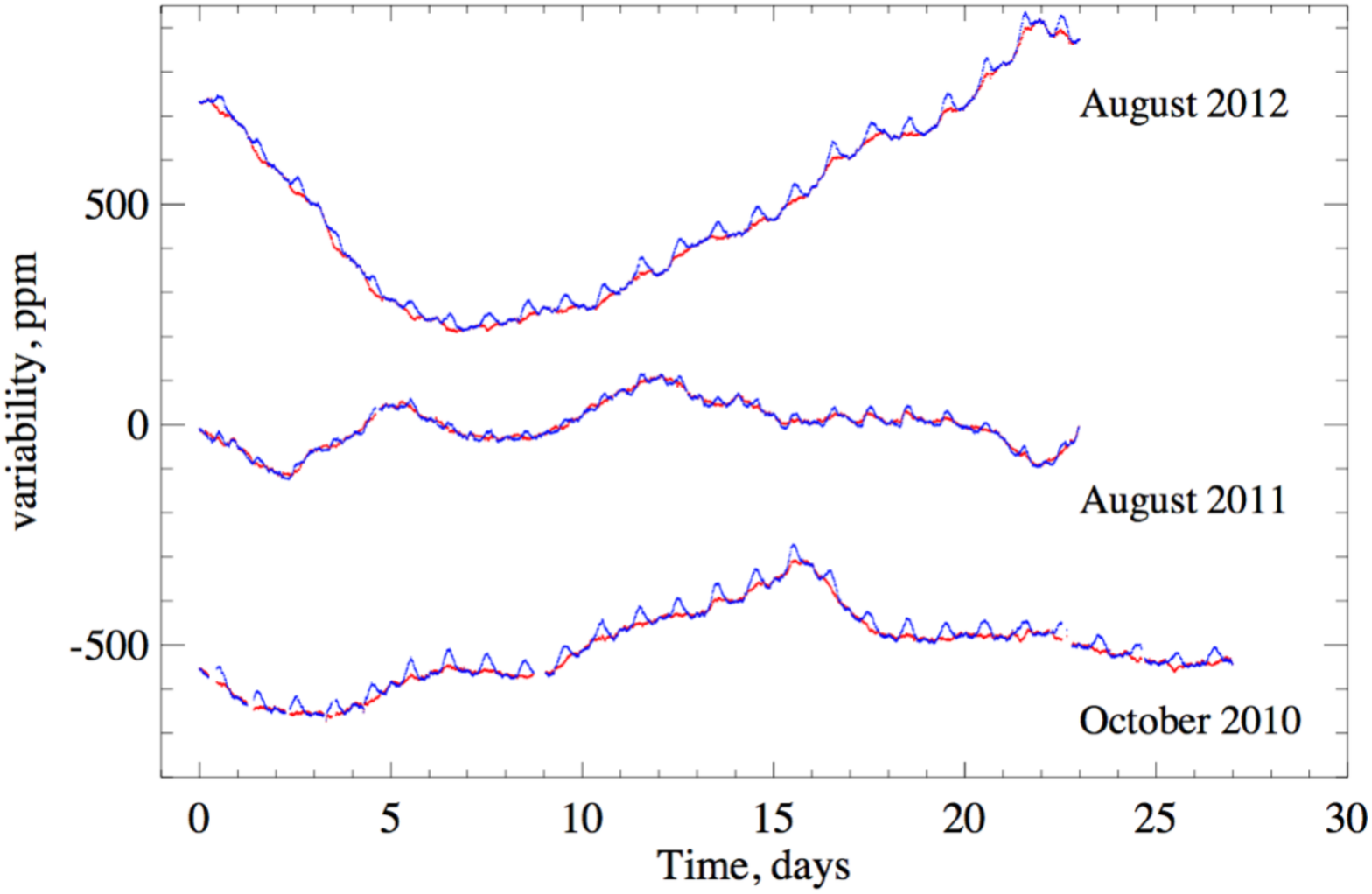}}
\caption{{\bf Correction of the facular TSI component for the instrumental 24-hour harmonic.} The facular TSI component calculated utilising 12-minute HMI/SDO magnetograms before (blue) and after (red) correction for the instrumental 24-hour harmonic.}
\label{fig:mod_sp}
\end{figure}

\begin{figure}
\resizebox{0.75\hsize}{!}{\includegraphics{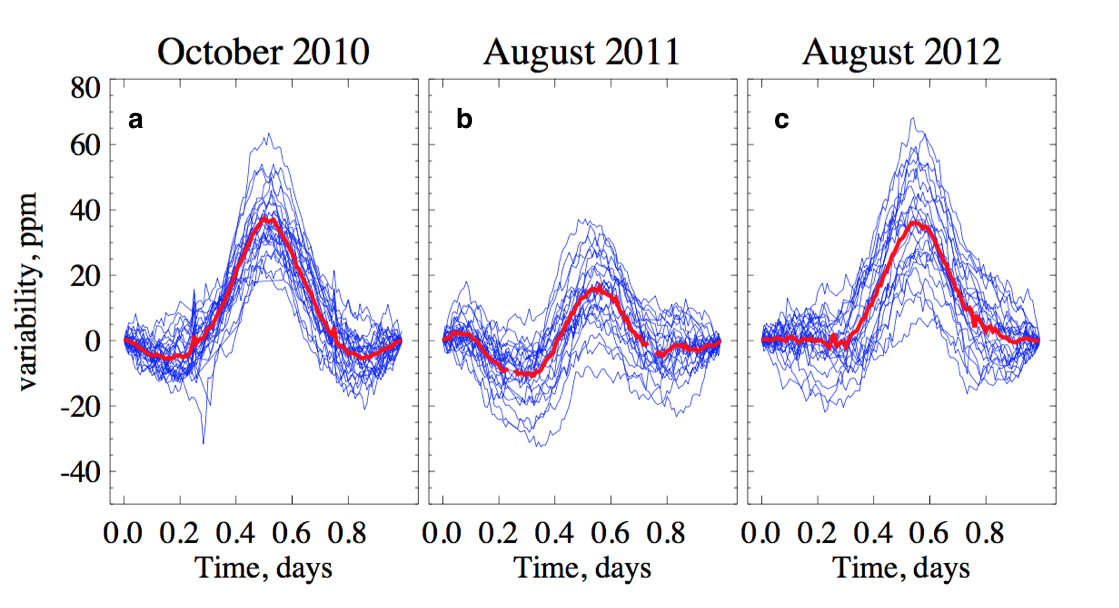}}
\caption{{\bf The 24-hour harmonic in HMI/SDO magnetograms. }The facular TSI component calculated utilising 12-minute HMI/SDO magnetograms is split into 24-hour intervals. Its variations in each interval are detrended and plotted over each other (blue curves). Red curves represent the averages of blue curves in each of the panels and are used to correct the TSI variations for the instrumental 24-hour harmonic.
}
\label{fig:mod_sp}
\end{figure}

\section*{Free parameter}\label{sub:free}
The SATIRE-S model has a single free parameter $B_{\rm sat}$, which represents the minimum magnetogram signal from a pixel fully filled by the magnetic field\cite{SATIRE}. If the magnetic signal in the pixel, $B$, is smaller than  the $B_{\rm sat}$  value, then it is assumed that $B/B_{\rm sat}$ part of the pixel is covered by faculae. If $B>B_{\rm sat}$ then the pixel is assumed to be fully covered by faculae. {\it Yeo et al. 2014}\cite{yeoetal2014}  found 
that the $B_{\rm sat}$ value of 230 G optimises the agreement between the TSI output of SATIRE-S and TSI time series obtained by the PMO6V radiometer on SoHO/VIRGO. 

To determine the $B_{\rm sat}$ value for 12-minute magnetograms we downloaded one SDO/HMI magnetogram per day for 30 April 2010 -- 31 July 2015 period and calculated daily TSI reconstruction. A $B_{\rm sat}$ value of 280 G was found by optimising the agreement between our TSI reconstruction and that of {\it Yeo et al. 2014}\cite{yeoetal2014}. This value is greater than that in {\it Yeo et al. 2014}\cite{yeoetal2014} because of the presence of a larger amount of small-scale weak magnetic features in magnetograms employed in our study.



\section*{Granulation-driven TSI variability}\label{sub:conv}
To quantify the variability introduced by convective motions at the solar surface we used  purely hydrodynamic (i.e. with no magnetic field) simulations with the MURAM\cite{MURAM} (stands for MPS/University of Chicago Radiative MHD) code.
These simulations cover the upper part of the convection zone and the photosphere in cartesian boxes. To use
these boxes to quantify brightness variability we would ideally tile the solar surface in the way describe by {\it Beeck et al. 2013a}\cite{beeck1}
using a relatively long simulation covering multiple supergranules. We would then perform detailed calculations of the
radiative transfer at a large number of viewing angles and then combine these to obtain the strength of the fluctuations.

To test whether this full apparatus was required, we performed this  detailed analysis for the small
(9~Mm in both horizontal directions, 3~Mm in the vertical direction) G2V solar simulations
described in {\it Beeck et al. 2013b}\cite{beeck2}. To save time for the purposes of the test, we used four snapshots separated in time by
19 minutes (making the radiative output essentially uncorrelated for these small boxes as the granulation structure has changed completely between any two snapshots). For each position the mean and standard
deviation were calculated and these were used to calculate the disk integrated average brightness and variability under the assumption
that the fluctuations at different positions were uncorrelated. We found that the relative variability (here defined as the standard deviation
divided by the mean) was $2.05\times 10^{-5}$. We then calculated the relative variability using the same four snapshots assuming that the
variability of the flux per unit area on the disk is independent of disk position. This yielded a variability of $1.88\times 10^{-5}$
which is within 8\% of the more detailed calculation.

On the basis of the above finding we then used the vertically emerging radiative flux from a time sequence from a large hydrodynamic  simulation
(196~Mm in the horizontal directions). The variability at different timescales was derived, and this was converted to a variability
for the entire disk of the Sun by scaling it by a factor of $\sqrt{d^2/\pi R_\odot^2}$, where $d=196$~Mm.

\section*{Comparison of calculated and observed TSI variations}\label{sub:TS}
Using high-cadence full-disk observations from SDO/HMI we reconstructed magnetically-driven TSI variability over the three one-month intervals considered in this study (see Table~1) with 12-minute cadence. The granulation-driven TSI variability, at 1-minute cadence, was simulated over the examined time intervals by consecutively repeating the 15-hour MURAM time series. This was imposed on the magnetically-driven TSI variability to yield 1-minute cadence time series of magnetically- and granulation-driven TSI variability.


We compared the TSI reconstruction to the 2-minute cadence TSI measurements obtained by PICARD/PREMOS and the 1-minute cadence TSI measurements from SOHO/VIRGO. Since variations on the 2-minute timescale (which corresponds to the Nyquist frequency of 1-minute cadence data) are still dominated by the p-mode oscillations and PREMOS data are only available with 2-minute cadence we considered every second point in the synthesised and VIRGO time series.

There are gaps in the VIRGO, PREMOS, and modelled time series (the latter are caused by breaks in the SDO/HMI data). In each comparison, we only considered the times at which data from  {\it all} considered time series are available. For example, the wavelet power spectra in Fig. 1c, f, i have been calculated utilising only those times at which VIRGO, PREMOS, and modelled data points were {\it simultaneously} present (this is the reason for orange curves, representing power spectrum of granulation component of TSI variability, being not identical in Fig. 1b,e,h).  The proportion of the period for which all three sources of data are available is about 90\% in all three considered time intervals. The gaps have been filled by linearly interpolating across them.

The daily TSI values for the  ``Long'' interval (see Table~1) have been calculated following the approach of  {\it Yeo et al. 2014}\cite{yeoetal2014} and compared to the observed values from the PMOD composite\cite{pmod_comp}.  The latter is currently a de facto standard observational dataset and is largely based on the VIRGO data for the period considered in this study. The modelled and measured power spectra shown in Fig.~2 were obtained by merging 4 minutes -- 7 days  and 7 days -- 19 years  parts of corresponding power spectra calculated for the August 2011 and ``Long''  intervals (see Table 1).  

\end{methods}

\begin{addendum}
 \item[Contributions] A.I.S., S.K.S., and N.A.K. designed the study. A.I.S. performed the calculations with contributions from R.H.C. who provided MURAM time series and K.L.Y. who prepared HMI/SDO magnetograms and corrected them for noise. N.A.K. and S.K.S. have led the development of the SATIRE code, R.H.C. actively contributed to the development of the MURAM code used in this study. W.K.S. provided PREMOS data and expertise on the TSI data. A.I.S., S.K.S., N.A.K., R.H.C. wrote the paper. 
 \item The research leading to this paper has received funding from the People Programme (Marie Curie Actions) of the European Union's Seventh Framework Programme (FP7/2007-2013) under REA grant agreement No. 624817 and European Research Council (ERC) under the European Union's Horizon 2020 research and innovation programme (grant agreement No. 715947). It also got financial support  from the BK21 plus program through the National Research Foundation (NRF) funded by the Ministry of Education of Korea.
 \item[Competing Interests] The authors declare that they have no
competing financial interests.
 \item[Correspondence] Correspondence and requests for materials
should be addressed to \\ A.I.S.~(email: shapiroa@mps.mpg.de).
\end{addendum}


\begin{thebibliography}{10}
\expandafter\ifx\csname url\endcsname\relax
  \def\url#1{\texttt{#1}}\fi
\expandafter\ifx\csname urlprefix\endcsname\relax\def\urlprefix{URL }\fi
\providecommand{\bibinfo}[2]{#2}
\providecommand{\eprint}[2][]{\url{#2}}

\bibitem{Claus_rev}
\bibinfo{author}{{Fr{\"o}hlich}, C.}
\newblock \bibinfo{title}{{Total Solar Irradiance: What Have We Learned from
  the Last Three Cycles and the Recent Minimum?}}
\newblock \emph{\bibinfo{journal}{Space Sci. Rev.}}
  \textbf{\bibinfo{volume}{176}}, \bibinfo{pages}{237--252}
  (\bibinfo{year}{2013}).

\bibitem{Greg2016}
\bibinfo{author}{{Kopp}, G.}
\newblock \bibinfo{title}{{Magnitudes and timescales of total solar irradiance
  variability}}.
\newblock \emph{\bibinfo{journal}{J. Space Weather and Space Clim.}}
  \textbf{\bibinfo{volume}{6}}, \bibinfo{pages}{A30} (\bibinfo{year}{2016}).

\bibitem{noise_limit}
\bibinfo{author}{{Rabello-Soares}, M.~C.}, \bibinfo{author}{{Roca Cortes}, T.},
  \bibinfo{author}{{Jimenez}, A.}, \bibinfo{author}{{Andersen}, B.~N.} \&
  \bibinfo{author}{{Appourchaux}, T.}
\newblock \bibinfo{title}{{An estimate of the solar background irradiance power
  spectrum.}}
\newblock \emph{\bibinfo{journal}{\aap}} \textbf{\bibinfo{volume}{318}},
  \bibinfo{pages}{970--974} (\bibinfo{year}{1997}).

\bibitem{TOSCA2012}
\bibinfo{author}{{Ermolli}, I.} \emph{et~al.}
\newblock \bibinfo{title}{{Recent variability of the solar spectral irradiance
  and its impact on climate modelling}}.
\newblock \emph{\bibinfo{journal}{Atmosph. Chem. Phys.}}
  \textbf{\bibinfo{volume}{13}}, \bibinfo{pages}{3945--3977}
  (\bibinfo{year}{2013}).

\bibitem{MPS_AA}
\bibinfo{author}{{Solanki}, S.~K.}, \bibinfo{author}{{Krivova}, N.~A.} \&
  \bibinfo{author}{{Haigh}, J.~D.}
\newblock \bibinfo{title}{{Solar Irradiance Variability and Climate}}.
\newblock \emph{\bibinfo{journal}{Ann. Rev. Astron. Astrophys.}}
  \textbf{\bibinfo{volume}{51}}, \bibinfo{pages}{311--351}
  (\bibinfo{year}{2013}).

\bibitem{Shapiro2014_stars}
\bibinfo{author}{{Shapiro}, A.~I.} \emph{et~al.}
\newblock \bibinfo{title}{{Variability of Sun-like stars: reproducing observed
  photometric trends}}.
\newblock \emph{\bibinfo{journal}{\aap}} \textbf{\bibinfo{volume}{569}},
  \bibinfo{pages}{A38} (\bibinfo{year}{2014}).

\bibitem{meunieretal2015}
\bibinfo{author}{{Meunier}, N.}, \bibinfo{author}{{Lagrange}, A.-M.},
  \bibinfo{author}{{Borgniet}, S.} \& \bibinfo{author}{{Rieutord}, M.}
\newblock \bibinfo{title}{{Using the Sun to estimate Earth-like planet
  detection capabilities. VI. Simulation of granulation and supergranulation
  radial velocity and photometric time series}}.
\newblock \emph{\bibinfo{journal}{\aap}} \textbf{\bibinfo{volume}{583}},
  \bibinfo{pages}{A118} (\bibinfo{year}{2015}).

\bibitem{HMI_12}
\bibinfo{author}{{Hoeksema}, J.~T.} \emph{et~al.}
\newblock \bibinfo{title}{{The Helioseismic and Magnetic Imager (HMI) Vector
  Magnetic Field Pipeline: Overview and Performance}}.
\newblock \emph{\bibinfo{journal}{\solphys}} \textbf{\bibinfo{volume}{289}},
  \bibinfo{pages}{3483--3530} (\bibinfo{year}{2014}).

\bibitem{yeoetal2014}
\bibinfo{author}{{Yeo}, K.~L.}, \bibinfo{author}{{Krivova}, N.~A.},
  \bibinfo{author}{{Solanki}, S.~K.} \& \bibinfo{author}{{Glassmeier}, K.~H.}
\newblock \bibinfo{title}{{Reconstruction of total and spectral solar
  irradiance from 1974 to 2013 based on KPVT, SoHO/MDI, and SDO/HMI
  observations}}.
\newblock \emph{\bibinfo{journal}{\aap}} \textbf{\bibinfo{volume}{570}},
  \bibinfo{pages}{A85} (\bibinfo{year}{2014}).

\bibitem{krivovaetal2003}
\bibinfo{author}{{Krivova}, N.~A.}, \bibinfo{author}{{Solanki}, S.~K.},
  \bibinfo{author}{{Fligge}, M.} \& \bibinfo{author}{{Unruh}, Y.~C.}
\newblock \bibinfo{title}{{Reconstruction of solar irradiance variations in
  cycle 23: Is solar surface magnetism the cause?}}
\newblock \emph{\bibinfo{journal}{\aap}} \textbf{\bibinfo{volume}{399}},
  \bibinfo{pages}{L1--L4} (\bibinfo{year}{2003}).

\bibitem{SATIRE}
\bibinfo{author}{{Krivova}, N.~A.}, \bibinfo{author}{{Solanki}, S.~K.} \&
  \bibinfo{author}{{Unruh}, Y.~C.}
\newblock \bibinfo{title}{{Towards a long-term record of solar total and
  spectral irradiance}}.
\newblock \emph{\bibinfo{journal}{J. Atmos. Solar-Terr. Phys.}}
  \textbf{\bibinfo{volume}{73}}, \bibinfo{pages}{223--234}
  (\bibinfo{year}{2011}).

\bibitem{MURAM}
\bibinfo{author}{{V{\"o}gler}, A.} \emph{et~al.}
\newblock \bibinfo{title}{{Simulations of magneto-convection in the solar
  photosphere. Equations, methods, and results of the MURaM code}}.
\newblock \emph{\bibinfo{journal}{\aap}} \textbf{\bibinfo{volume}{429}},
  \bibinfo{pages}{335--351} (\bibinfo{year}{2005}).

\bibitem{COROT2}
\bibinfo{author}{{Bord{\'e}}, P.}, \bibinfo{author}{{Rouan}, D.} \&
  \bibinfo{author}{{L{\'e}ger}, A.}
\newblock \bibinfo{title}{{Exoplanet detection capability of the COROT space
  mission}}.
\newblock \emph{\bibinfo{journal}{\aap}} \textbf{\bibinfo{volume}{405}},
  \bibinfo{pages}{1137--1144} (\bibinfo{year}{2003}).

\bibitem{KEPLER}
\bibinfo{author}{{Borucki}, W.~J.} \emph{et~al.}
\newblock \bibinfo{title}{{Kepler Planet-Detection Mission: Introduction and
  First Results}}.
\newblock \emph{\bibinfo{journal}{Science}} \textbf{\bibinfo{volume}{327}},
  \bibinfo{pages}{977--} (\bibinfo{year}{2010}).

\bibitem{basrietal2013}
\bibinfo{author}{{Basri}, G.}, \bibinfo{author}{{Walkowicz}, L.~M.} \&
  \bibinfo{author}{{Reiners}, A.}
\newblock \bibinfo{title}{{Comparison of Kepler Photometric Variability with
  the Sun on Different Timescales}}.
\newblock \emph{\bibinfo{journal}{\apj}} \textbf{\bibinfo{volume}{769}},
  \bibinfo{pages}{37} (\bibinfo{year}{2013}).

\bibitem{TESS}
\bibinfo{author}{{Ricker}, G.~R.} \emph{et~al.}
\newblock \bibinfo{title}{{Transiting Exoplanet Survey Satellite (TESS)}}.
\newblock \emph{\bibinfo{journal}{Journal of Astronomical Telescopes,
  Instruments, and Systems}} \textbf{\bibinfo{volume}{1}},
  \bibinfo{pages}{014003} (\bibinfo{year}{2015}).

\bibitem{PLATO}
\bibinfo{author}{{Rauer}, H.} \emph{et~al.}
\newblock \bibinfo{title}{{The PLATO 2.0 mission}}.
\newblock \emph{\bibinfo{journal}{Exper. Astron.}}
  \textbf{\bibinfo{volume}{38}}, \bibinfo{pages}{249--330}
  (\bibinfo{year}{2014}).

\bibitem{rev_mag}
\bibinfo{author}{{Solanki}, S.~K.}, \bibinfo{author}{{Inhester}, B.} \&
  \bibinfo{author}{{Sch{\"u}ssler}, M.}
\newblock \bibinfo{title}{{The solar magnetic field}}.
\newblock \emph{\bibinfo{journal}{Rep. Progr. Phys.}}
  \textbf{\bibinfo{volume}{69}}, \bibinfo{pages}{563--668}
  (\bibinfo{year}{2006}).

\bibitem{HMI}
\bibinfo{author}{{Schou}, J.} \emph{et~al.}
\newblock \bibinfo{title}{{Design and Ground Calibration of the Helioseismic
  and Magnetic Imager (HMI) Instrument on the Solar Dynamics Observatory
  (SDO)}}.
\newblock \emph{\bibinfo{journal}{\solphys}} \textbf{\bibinfo{volume}{275}},
  \bibinfo{pages}{229--259} (\bibinfo{year}{2012}).

\bibitem{beeck2}
\bibinfo{author}{{Beeck}, B.}, \bibinfo{author}{{Cameron}, R.~H.},
  \bibinfo{author}{{Reiners}, A.} \& \bibinfo{author}{{Sch{\"u}ssler}, M.}
\newblock \bibinfo{title}{{Three-dimensional simulations of near-surface
  convection in main-sequence stars. II. Properties of granulation and spectral
  lines}}.
\newblock \emph{\bibinfo{journal}{\aap}} \textbf{\bibinfo{volume}{558}},
  \bibinfo{pages}{A49} (\bibinfo{year}{2013}).

\bibitem{aigrain2004}
\bibinfo{author}{{Aigrain}, S.}, \bibinfo{author}{{Favata}, F.} \&
  \bibinfo{author}{{Gilmore}, G.}
\newblock \bibinfo{title}{{Characterising stellar micro-variability for
  planetary transit searches}}.
\newblock \emph{\bibinfo{journal}{\aap}} \textbf{\bibinfo{volume}{414}},
  \bibinfo{pages}{1139--1152} (\bibinfo{year}{2004}).

\bibitem{Penza2009}
\bibinfo{author}{Penza, V.} \& \bibinfo{author}{del Moro, D.}
\newblock \bibinfo{title}{{A granulation model: possible effects of contrast
  variations on the solar irradiance along the cycle}}.
\newblock \emph{\bibinfo{journal}{Mem. Soc. Astron. It.}}
  (\bibinfo{year}{2009}).

\bibitem{rast2003}
\bibinfo{author}{{Rast}, M.~P.}
\newblock \bibinfo{title}{{The Scales of Granulation, Mesogranulation, and
  Supergranulation}}.
\newblock \emph{\bibinfo{journal}{\apj}} \textbf{\bibinfo{volume}{597}},
  \bibinfo{pages}{1200--1210} (\bibinfo{year}{2003}).

\bibitem{seleznyovetal2011}
\bibinfo{author}{{Seleznyov}, A.~D.}, \bibinfo{author}{{Solanki}, S.~K.} \&
  \bibinfo{author}{{Krivova}, N.~A.}
\newblock \bibinfo{title}{{Modelling solar irradiance variability on time
  scales from minutes to months}}.
\newblock \emph{\bibinfo{journal}{\aap}} \textbf{\bibinfo{volume}{532}},
  \bibinfo{pages}{A108} (\bibinfo{year}{2011}).

\bibitem{werner_PREMOS}
\bibinfo{author}{Schmutz, W.}, \bibinfo{author}{Fehlmann, A.},
  \bibinfo{author}{Finsterle, W.}, \bibinfo{author}{Kopp, G.} \&
  \bibinfo{author}{Thuillier, G.}
\newblock \bibinfo{title}{{Total solar irradiance measurements with
  PREMOS/PICARD}}.
\newblock In \emph{\bibinfo{booktitle}{Proc. Int. Rad. Symp.}},
  \bibinfo{pages}{624--627} (\bibinfo{year}{2016}).

\bibitem{VIRGO}
\bibinfo{author}{{Fr{\"o}hlich}, C.} \emph{et~al.}
\newblock \bibinfo{title}{{VIRGO: Experiment for Helioseismology and Solar
  Irradiance Monitoring}}.
\newblock \emph{\bibinfo{journal}{\solphys}} \textbf{\bibinfo{volume}{162}},
  \bibinfo{pages}{101--128} (\bibinfo{year}{1995}).

\bibitem{pmod_comp}
\bibinfo{author}{{Fr{\"o}hlich}, C.}
\newblock \bibinfo{title}{{Solar Irradiance Variability Since 1978. Revision of
  the PMOD Composite during Solar Cycle 21}}.
\newblock \emph{\bibinfo{journal}{Space Sci. Rev.}}
  \textbf{\bibinfo{volume}{125}}, \bibinfo{pages}{53--65}
  (\bibinfo{year}{2006}).

\bibitem{shapiroetal2016}
\bibinfo{author}{{Shapiro}, A.~I.}, \bibinfo{author}{{Solanki}, S.~K.},
  \bibinfo{author}{{Krivova}, N.~A.}, \bibinfo{author}{{Yeo}, K.~L.} \&
  \bibinfo{author}{{Schmutz}, W.~K.}
\newblock \bibinfo{title}{{Are solar brightness variations faculae- or
  spot-dominated?}}
\newblock \emph{\bibinfo{journal}{\aap}} \textbf{\bibinfo{volume}{589}},
  \bibinfo{pages}{A46} (\bibinfo{year}{2016}).

\bibitem{Aigrainetal2015}
\bibinfo{author}{Aigrain, S.} \emph{et~al.}
\newblock \bibinfo{title}{{Testing the recovery of stellar rotation signals
  from Kepler light curves using a blind hare-and-hounds exercise}}.
\newblock \emph{\bibinfo{journal}{Mon. Not. R. Astron. Soc.}}
  \textbf{\bibinfo{volume}{450}}, \bibinfo{pages}{3211--3226}
  (\bibinfo{year}{2015}).

\bibitem{Ludwiget2009}
\bibinfo{author}{{Ludwig}, H.-G.} \emph{et~al.}
\newblock \bibinfo{title}{{Hydrodynamical simulations of convection-related
  stellar micro-variability. II. The enigmatic granulation background of the
  CoRoT target HD 49933}}.
\newblock \emph{\bibinfo{journal}{\aap}} \textbf{\bibinfo{volume}{506}},
  \bibinfo{pages}{167--173} (\bibinfo{year}{2009}).

\bibitem{Claus1997}
\bibinfo{author}{Fr{\"o}hlich, C.} \emph{et~al.}
\newblock \bibinfo{title}{{First Results from VIRGO, the Experiment for
  Helioseismology and Solar Irradiance Monitoring on SOHO}}.
\newblock \emph{\bibinfo{journal}{Solar Physics}}
  \textbf{\bibinfo{volume}{170}}, \bibinfo{pages}{1--25}
  (\bibinfo{year}{1997}).

\bibitem{FandL2004}
\bibinfo{author}{Fr{\"o}hlich, C.} \& \bibinfo{author}{Lean, J.}
\newblock \bibinfo{title}{{Solar radiative output and its variability: Evidence
  and mechanisms}}.
\newblock \emph{\bibinfo{journal}{Astron. Astrophys. Rev.}}
  \textbf{\bibinfo{volume}{12}}, \bibinfo{pages}{273--320}
  (\bibinfo{year}{2004}).

\bibitem{sat_spectra}
\bibinfo{author}{{Unruh}, Y.~C.}, \bibinfo{author}{{Solanki}, S.~K.} \&
  \bibinfo{author}{{Fligge}, M.}
\newblock \bibinfo{title}{{The spectral dependence of facular contrast and
  solar irradiance variations}}.
\newblock \emph{\bibinfo{journal}{\aap}} \textbf{\bibinfo{volume}{345}},
  \bibinfo{pages}{635--642} (\bibinfo{year}{1999}).

\bibitem{kurucz1992}
\bibinfo{author}{{Kurucz}, R.~L.}
\newblock \bibinfo{title}{{''Finding'' the ''missing'' solar ultraviolet
  opacity.}}
\newblock \emph{\bibinfo{journal}{Rev. Mex. de Astron. Astrof., vol.~23}}
  \textbf{\bibinfo{volume}{23}}, \bibinfo{pages}{181--186}
  (\bibinfo{year}{1992}).

\bibitem{ATLAS9_CK}
\bibinfo{author}{{Castelli}, F.} \& \bibinfo{author}{{Kurucz}, R.~L.}
\newblock \bibinfo{title}{{Model atmospheres for VEGA}}.
\newblock \emph{\bibinfo{journal}{\aap}} \textbf{\bibinfo{volume}{281}},
  \bibinfo{pages}{817--832} (\bibinfo{year}{1994}).

\bibitem{balletal2014}
\bibinfo{author}{{Ball}, W.~T.}, \bibinfo{author}{{Krivova}, N.~A.},
  \bibinfo{author}{{Unruh}, Y.~C.}, \bibinfo{author}{{Haigh}, J.~D.} \&
  \bibinfo{author}{{Solanki}, S.~K.}
\newblock \bibinfo{title}{{A New SATIRE-S Spectral Solar Irradiance
  Reconstruction for Solar Cycles 21--23 and Its Implications for Stratospheric
  Ozone}}.
\newblock \emph{\bibinfo{journal}{J. Atmos. Sci.}}
  \textbf{\bibinfo{volume}{71}}, \bibinfo{pages}{4086--4101}
  (\bibinfo{year}{2014}).

\bibitem{ortizetal2002}
\bibinfo{author}{{Ortiz}, A.}, \bibinfo{author}{{Solanki}, S.~K.},
  \bibinfo{author}{{Domingo}, V.}, \bibinfo{author}{{Fligge}, M.} \&
  \bibinfo{author}{{Sanahuja}, B.}
\newblock \bibinfo{title}{{On the intensity contrast of solar photospheric
  faculae and network elements}}.
\newblock \emph{\bibinfo{journal}{\aap}} \textbf{\bibinfo{volume}{388}},
  \bibinfo{pages}{1036--1047} (\bibinfo{year}{2002}).

\bibitem{beeck1}
\bibinfo{author}{{Beeck}, B.}, \bibinfo{author}{{Cameron}, R.~H.},
  \bibinfo{author}{{Reiners}, A.} \& \bibinfo{author}{{Sch{\"u}ssler}, M.}
\newblock \bibinfo{title}{{Three-dimensional simulations of near-surface
  convection in main-sequence stars. I. Overall structure}}.
\newblock \emph{\bibinfo{journal}{\aap}} \textbf{\bibinfo{volume}{558}},
  \bibinfo{pages}{A48} (\bibinfo{year}{2013}).

\end{thebibliography}
\end{document}